\documentclass[a4paper,conference]{IEEEtran}
\ifCLASSINFOpdf
\else
\fi
\IEEEoverridecommandlockouts
\usepackage[colorlinks=true,linkcolor=black,urlcolor=black,citecolor=black]{hyperref}
\usepackage{cite}
\usepackage{amsmath,amssymb,amsfonts,bm}
\usepackage{algorithmic}
\usepackage{graphicx}
\usepackage{textcomp}
\usepackage{multirow}
\usepackage{subfigure}
\usepackage[dvipsnames]{xcolor}
\usepackage[bottom]{footmisc}

\begin{document}
%
\title{U-Former: Improving Monaural Speech Enhancement with Multi-head Self and Cross Attention}

\author{\IEEEauthorblockN{\textit{Xinmeng Xu$^{1}$, Jianjun Hao$^{2,*}$ \thanks{$^*$Corresponding author}}}
\IEEEauthorblockA{$^{1}$E.E. Engineering, Trinity College Dublin, Dublin, Ireland \\
$^{2}$School of Foreign Languages, Hubei University of Chinese Medicine, Wuhan, China\\
\texttt{xux3@tcd.ie, loyalcolin@163.com}}
}


%


\maketitle

\begin{abstract}
For supervised speech enhancement, contextual information is important for accurate spectral mapping. However, commonly used deep neural networks (DNNs) are limited in capturing temporal contexts. To leverage long-term contexts for tracking a target speaker, this paper treats the speech enhancement as sequence-to-sequence mapping, and propose a novel monaural speech enhancement U-net structure based on Transformer, dubbed U-Former. The key idea is to model long-term correlations and dependencies, which are crucial for accurate noisy speech modeling, through the multi-head attention mechanisms. For this purpose, U-Former incorporates multi-head attention mechanisms at two levels: 1) a multi-head self-attention module which calculate the attention map along both time- and frequency-axis to generate time and frequency sub-attention maps for leveraging global interactions between encoder features, while 2) multi-head cross-attention module which are inserted in the skip connections allows a fine recovery in the decoder by filtering out uncorrelated features. Experimental results illustrate that the U-Former obtains consistently better performance than recent models of PESQ, STOI, and SSNR scores. The source code is available at: \scriptsize{\href{https://github.com/XinmengXu/Uformer.git}{\color{RubineRed}{\textit{https://github.com/XinmengXu/Uformer.git}}}}. 
\end{abstract}


%
\IEEEpeerreviewmaketitle

\section{Introduction}
The problem of enhancing speech corrupted by uncorrelated additive noise, when only single-channel noisy speech is available, has been widely studied in the past and it is still a active filed of research. Traditional speech enhancement approaches include spectral subtraction \cite{ref1}, Wiener filtering \cite{ref2}, statistical model-based methods \cite{ref3}, and non-negative matrix factorization \cite{ref4}. Recently, supervised methods using deep neural networks have become shown their promising performance on monaural speech enhancement \cite{ref6, ref7}.

Various encoder-decoder frameworks based on convolutional neural networks (CNN) or recurrent neural networks (RNNs) have gained superior network performance in speech enhancement. For modeling long-range sequence like speech, CNN requires more convolutional layers to enlarge receptive field \cite{ref9, ref101}. The dilated convolutional neural network has been proposed for processing the long-term temporal sequence \cite{ref21}. In addition, recurrent neural networks (RNNs) employing long short-term memory (LSTM) have demonstrated a robust speech enhancement performance \cite{ref12}. However, RNNs require large number of parameters and lengthy training times, and the memory of LSTM is limited, making it prone to forgetting distant information \cite{ref14}. Temporal convolutional networks (TCNs) are proposed to match the speech enhancement performance of RNNs while use fewer parameters and requiring less time to train \cite{ref15}. TCNs utilize causal dilated kernels to garner a fixed-size receptive field over previous frames, allowing the modelling of long-term dependencies. However, the performance of TCNs are adversely affected when events occur out of expected order due to the positional nature of the kernels \cite{ref16}.

Recently, as the outstanding performance in natural language processing (NLP), the Transformer is introduced in speech enhancement problem \cite{ref17}. The Transformer follows the encoder and decoder architecture with stacked self-attention and point-wise feed-forward layers \cite{ref18}, which promise a  stronger long-term dependencies modelling capability than TCNs. However, the Transformer has a limitation as difficult to process a multiple-sequence alignment as an input for feature extraction in the inference stage \cite{ref19}. In this study, we integrate the Transformer and encoder-decoder structure for learning both local and global contextual information of long-range speech. Consequently, this paper propose a U-Former for monaural speech enhancement, which keeps the inductive bias of convolution by using a U-Net architecture, but introduces attention mechanisms at two levels, 1) the multi-head self-attention (MHSA) module calculates the attention map along both time- and frequency-axis to generate time and frequency sub-attention maps for leveraging global interactions between encoder feature, 2) the multi-head cross-attention module (MHCA) in skip-connections to filter out uncorrelated feature and to  reduce the loss of feature information from corresponding encoder layer \footnote{Speech samples are available at: \scriptsize{\href{https://xinmengxu.github.io/AVSE/MHACUnet.html}{\texttt{https://xinmengxu.github.io/\\AVSE/MHACUnet.html}}}}. .

\begin{figure*}[t]
  \centering
  \includegraphics[width=\linewidth]{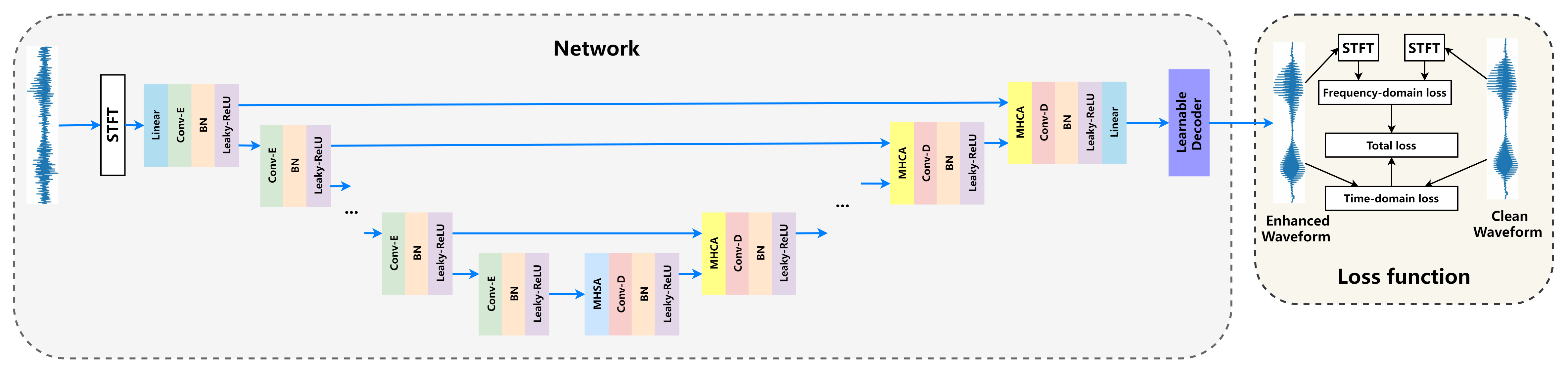}
  \caption{Network architecture of proposed U-Former. Conv-E, Conv-D, and BN denote convolution encoder, convolution decoder, and batch normalization respectively.}
  \label{fig:1}
\end{figure*}

In a nutshell, the contributions of our work are three-fold:
\begin{itemize}
    \item We integrate the Transformer and encoder-decoder structure, which aims to learn both local and global contextual information of long-range speech.
    \item We propose MHSA and MHCA modules to leverage global interactions between encoder feature and to filter out uncorrelated feature and to  reduce the loss of feature information from corresponding encoder layer.
    \item To evaluate the proposed U-Former, we perform comprehensive ablation studies on public available speech data. Results show that U-Former outperforms the state-of-the-art methods.
\end{itemize}

Rest of paper is organised as follows: Section II introduces the signal model and problem formulation. Section III presents the proposed network, and discusses mechanisms of MHSA and MHCA in detail. Section IV describes the datasets and network parameters. Section V demonstrates the results and analysis, and a conclusion is shown in Section VI.

\section{Signal model and problem formulation}\label{sec2}

Given a single-microphone mixture $y$, monaural SE aims to separate target speech $s$ from background noise $n$. A noisy mixture can be formulated as
\begin{equation}
    y[k] = s[k] + n[k],
\label{eq1}
\end{equation}
where $k$ is the time sample index. Taking the STFT on both sides, we obtain
\begin{equation}
    Y(m,f)=S(m,f)+N(m,f),
\label{eq2}
\end{equation}
where $Y$, $S$ and $N$ represent the STFT of $y$, $s$ and $n$, respectively, and $m$ and $f$ index the time frame and frequency bin, respectively. The STFT of a speech signal can be expressed in Cartesian coordinates. Hence, Eq~\ref{eq2} can be rewritten into
\begin{equation}
\begin{aligned}
   Y^{(r)}(m,f)+iY^{(i)}(m,f)&=\Big(S^{(r)}(m,f)+N{(r)}(m,f)\Big)\\ 
   &+ i\Big(S^{(i)}(m,f)+N^{(i)}(m,f)\Big),   
\end{aligned}
\label{eq3}
\end{equation}
where the superscripts $(r)$ and $(i)$ indicate real and imaginary components, respectively. In our work, we refer the training target used in complex spectral mapping, i.e. $S^{(r)}$ and $S^{(i)}$, as the target RI-spectrogram. 

The speech enhancement using U-Former can be formulated as
\begin{equation}
    s_{pre}=f_{\theta}(Y^{(r)}, Y^{(i)})
\end{equation}
where $f_{\theta}$ denotes a function defining a U-Former model parameterized by $\theta$.

\section{Model Architecture}\label{sec3}
Fig \ref{fig:1} shows the proposed speech enhancement network. The model uses encoder-decoder structure that based on the standard convolution U-Net \cite{ref18}, as the main body. The input to the U-Former is complex valued spectrogram computed by short-time Fourier transform (STFT), denoted by $\textbf{Y}_{r,i}\in \mathbb{R}^{T\times F\times2}$, where $T$ represents the number of time steps and $F$ represents the number of frequency bands. The output of U-Former $\textbf{S}_{out}$ is processed by a learnable decoder layer to transform the feature back to the enhanced time-domain signal $\textbf{s}_{pre} \in \mathbb{R}$. The network structure of the learnable decoder is designed to be exactly the same as the convolutional implementation of inverse STFT. Specifically, the learnable decoder is a 1D transposed convolution layer whose kernel size and stride being the window length and hop size in the STFT \cite{ref22, ref23}. .

In addition, the encoder of U-Former contains several stacked convolutional layers, each of which is followed by batch normalization, and Leaky-ReLU for non-linearity. The decoder is mirror representations of the encoder. What is more, the proposed U-Former models long-range contextual interaction and dependencies by using two types of attention modules, MHSA placed in bottleneck part and MHCA placed in skip-connection part. Both modules are designed to express a new representation of the input based on its self-attention in the first case or on the attention paid to higher level features in the second.

\begin{figure}
  \centering
  \includegraphics[width=0.75\linewidth]{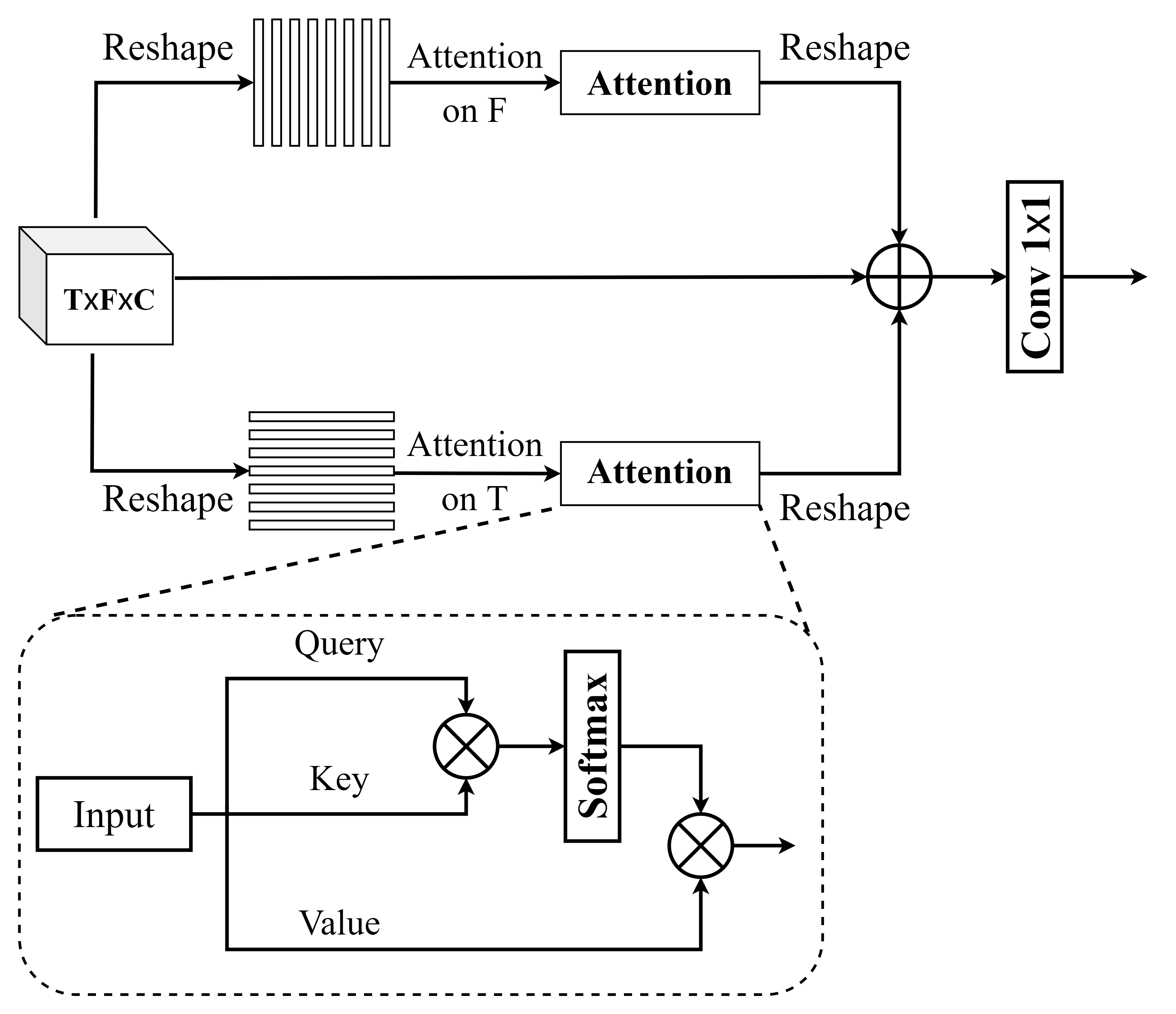}
  \caption{Schematic diagram of MHSA module.}
  \label{fig:2}
\end{figure}

\subsection{Multi-head Self-attention (MHSA)}

In the proposed U-Former, the MHSA module is designed to extract long range structural information \cite{ref17}. To this end, it is composed of multi-head self-attention functions positioned at the bottom of the U-Net as shown in Fig~\ref{fig:1}. MHSA takes as input an L-length sequence of embedding and produces a same size output sequence. The output of the attention matrix is presented as \cite{ref24}:
\begin{equation}
    \mathcal{A}(Q,K,V)=\text{Softmax}\Big(\frac{QK^{\top}}{\sqrt{L}}\Big)V,
\end{equation}
where $\top$ denotes the transpose symbol. Initially, the full attention map $\textbf{F}_{in}\in \mathbb{R^{T\times F\times C}}$ is the input to each sub-layer of the encoder, the proposed MHSA are reshaped the 2D attention map into 1D sub-maps in time- and frequency-axis as $T\times C_{\text{layer}}$ and $F\times C_{\text{layer}}$, respectively, as shown in Fig~\ref{fig:2}. Each sub attention map progrates information along specific axis. Moreover, two sub attention maps are constructed as parallel calculations on multiple GPUs to optimize the training process. The multi-head attention for the time direction can be presented as:
\begin{equation}
    \text{multihead}(Q^t, K^t, V^t)=[h_1; h_2;...;h_n]W_t^O,
\end{equation}
where
\begin{equation}
    h_i = \mathcal{A}\Big(Q^tW^Q_t, K^tW^K_t, V^tW^V_t\Big),
\end{equation}
where $W_t^O\in\mathbb{R}^{n\times C_v\times C_{\text{layer}}}$ ($n$ denotes the number of head), ${W^Q_t, W^K_t}\in\mathbb{R}^{C_{\text{layer}}\times C_k}$, $ W^V_t\in\mathbb{R}^{C_{\text{layer}}\times C_v}$. In the time attention map, query, keys and values are presented as $Q^t, K^t, V^t$, respectively. Similarly, for frequency attention, we adopt same equations but different notations $f$:
\begin{equation}
    \text{multihead}(Q^f, K^f, V^f)=[h_1; h_2;...;h_n]W_f^O,
\end{equation}
where
\begin{equation}
    h_i = \mathcal{A}\Big(Q^fW^Q_f, K^fW^K_f, V^fW^V_f\Big),
\end{equation}

The proposed MHSA demonstrates high efficacy to learn the long-term dependencies. Inspired by the position-sensitivity of self-attention \cite{ref38}, the output $h_i$ is enabled to retrieve relative positions by pooling over query-key affinities $Q^t_p K^t_p$, the key-dependent positional bias term $K_c^t r^{K^t}_{c-p}$:

\begin{equation}
\begin{aligned}
  h_i = \sum_{c\in \mathcal{N}_{1\times n(c)}} \text{softmax} (Q^t_p K^t_c + Q^t_p r^{Q^t}_{c-p}
  \\ + K_c^t r^{K^t}_{c-p})(V_c-r^{V^t}_{c-p}),
\end{aligned}
\end{equation}
where $\mathcal{N}_{1\times n(c)}$ is the local $1\times m$ region around position $c$. The learnable $r^{Q^t}_{c-p}$, $r^{K^t}_{c-p}$, and $r^{V^t}_{c-p}$ are the positional encoding for query, key, and value, respectively.

To capture global information, we employ two sub-attention layers consecutively for the time-axis and frequency-axis, respectively. Consequently, the outputs of time and frequency attentions can be written as:
\begin{equation}
\begin{aligned}
   M_t&=\text{multihead}(Q^t, K^t, V^t),\\
   M_f&=\text{multihead}(Q^f, K^f, V^f),
\end{aligned}
\end{equation}

Finally, the masked attention maps are reshaped, integrated and processed by a $1\times1$ convolution block to obtained the output feature map, $\textbf{F}_{out}$, in which residual connections are added as well.
\begin{equation}
    \textbf{F}_{out} = \text{Conv}(M_t+M_f+M_p),
\end{equation}
where $M_p$ is the input feature map.

\subsection{Multi-head Cross-attention (MHCA)}

The MHSA module allows to connect every element in the input with each other. Attention may also be used to increase the U-Net decoder efficiency and in particular enhance the lower level feature maps that are passed through the skip connections.

\begin{figure}
  \centering
  \includegraphics[width=0.75\linewidth]{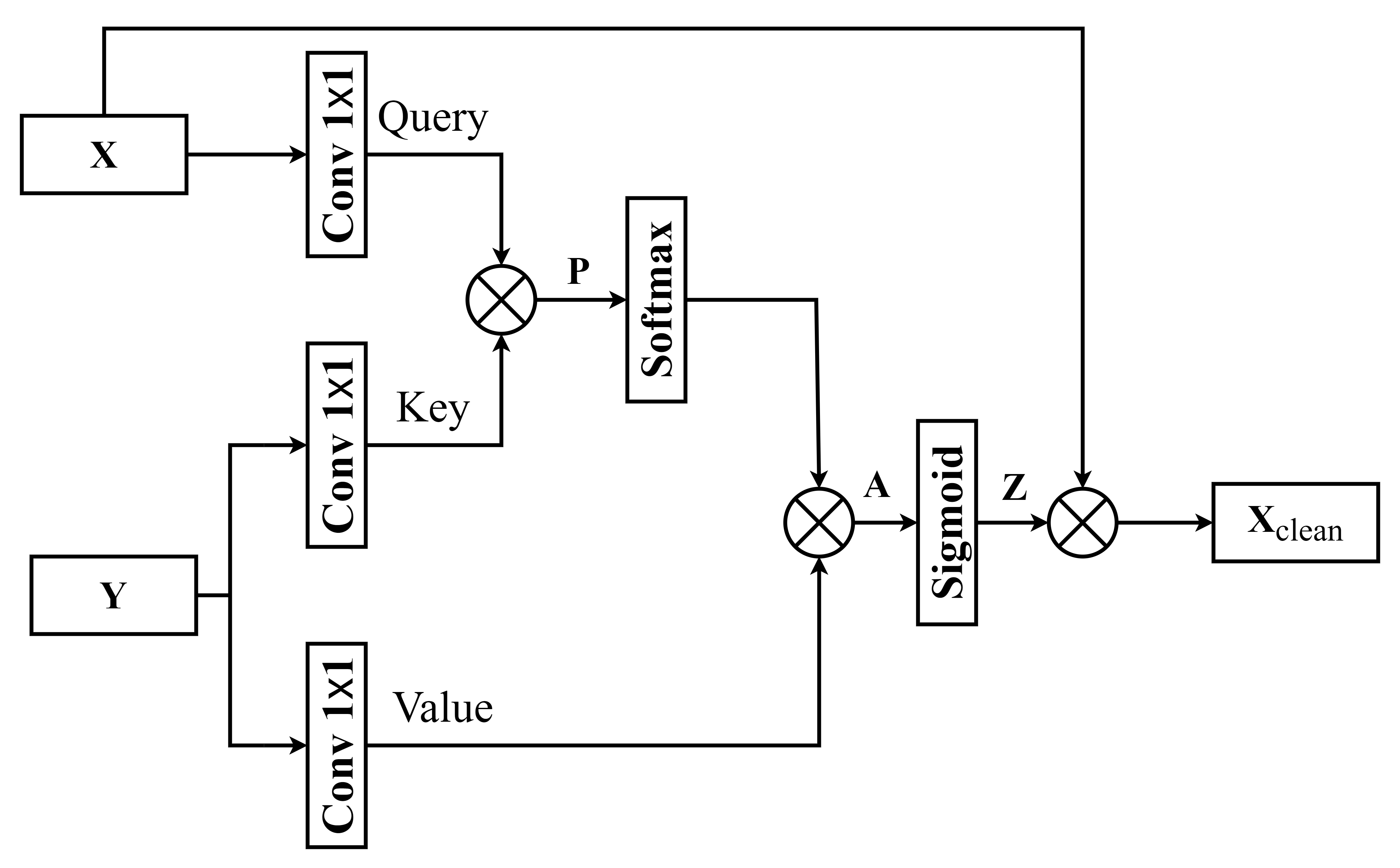}
  \caption{Schematic diagram of MHCA module.}
  \label{fig:3}
\end{figure}

The idea behind the MHCA module is to turn off irrelevant or noisy areas from the skip connection features and highlight regions that present a significant interest for the application. The MHCA block which diagram is shown in Fig~\ref{fig:3}, is designed as a gating operation of the skip connection.

The MHCA takes transformed feature maps corresponding to the decoder layer $\textbf{X}$, and processed by a $1\times1$ convolution block to form the query, and takes transformed feature maps corresponding to the encoder layer $\textbf{Y}$, to form the key-value pair by using two $1\times1$ convolution blocks. The MHCA first computes the weight $\textbf{P}$ for getting the attention component $\textbf{A},$ given by
\begin{equation}
    \textbf{P} = \frac{\textbf{Q}\textbf{K}^\top}{\sqrt{F}},
\end{equation}
where $\textbf{Q}$, $\textbf{K}$ are represented query and key, respectively, and $F$ denotes the number of frequency bins. Intuitively, the multiplication operation between $\textbf{Q}$ and $\textbf{K}$ emphasizes the regions which are slowly varying in time and have high power. What is more, the softmax function is used to generate attention mask $\widehat{\textbf{W}}  = \{\widehat{W}_{i,j}\}$,
\begin{equation}
  \widehat{W}_{i,j}  = \frac{exp(W_{i,j})}{w_j},
\end{equation}
where
\begin{equation}
  w_j = \sum_{i=1}^{F} exp(W_{i,j}),
\end{equation}

The attention component \textbf{A} is determined by 
\begin{equation}
  \textbf{A}=\widehat{\textbf{W}}  \textbf{V},
\end{equation}
where \textbf{V} denotes the index of value. What is more, the output of MHCA \textbf{Z} is computed by
\begin{equation}
  \textbf{Z} = \text{sigmoid}(\textbf{A}),
\end{equation}
Consequently, the \textbf{Z} is weight value that are re-scaled between 0 and 1 through a sigmoid activation function, in which low magnitude elements indicated irrelevant areas to be reduced. 

A cleaned up version of $ \textbf{X}_{clean}$ is given by
\begin{equation}
    \textbf{X}_{clean} = \textbf{X} \odot \textbf{Z},
\end{equation}
where $\odot$ is Hadamard product. Finally the $\textbf{X}_{clean}$ is concatenated with $\textbf{Y}$.

\subsection{Loss function}
\label{sssec:subsubhead}
The proposed model is trained by a combination of two losses. First, the time-domain loss that is defined as:
\begin{equation}
    L_t(\textbf{x}, \hat{\textbf{x}}) = \frac{1}{M}\sum_{n=0}^{M-1}(x_i[n]-\hat{x}_i[n])^2,
\end{equation}
where $x[n]$ and $\hat{x}[n]$ represent the $n^{th}$ sample of the clean and the enhanced utterance respectively and M is the utterance length.

Second, the frequency-domain loss which takes STFT of the utterances and use $L_1$ over the $L_1$ norm of the STFT coefficients \cite{ref20}, which is given by:
\begin{equation}
\begin{split}
    L_f(\textbf{x}, \hat{\textbf{x}})=\frac{1}{T\times F}\sum_{t=1}^T\sum_{f=1}^{F}|(|X(t,f)_r|+|X(t,f)_i|)\\
    -|(\hat{X}(t,f)_r|+|\hat{X}(t,f)_i|)|,
\end{split}
\end{equation}
where $X(t,f)$ and $\hat{X}(t,f)$ are the units of STFTs of $\textbf{x}$ and $\hat{\textbf{x}}$ respectively.$T$ and $F$ denote the index of frames and frequency bins. $X_r$ and $X_i$ represent real and the imaginary of complex variable $X$.

Finally, the time-frequency domain loss is expressed as:
\begin{equation}
     L(\textbf{x}, \hat{\textbf{x}})=\alpha\times L_t(\textbf{x}, \hat{\textbf{x}})+(1-\alpha)\times L_f(\textbf{x}, \hat{\textbf{x}}),
\end{equation}
where $\alpha$ is a hyper-parameter that is tuned on the validation set. The diagram of the loss function is shown in the right part of Fig~\ref{fig:1}.

\section{Experimental Setup}

\subsection{Datasets}
In order to evaluate performance of the proposed model, experiments are conducted on Librispeech corpus \cite{ref25}. There 6500 clean utterances are selected for training set and 400 are selected for validation set, which are created under the randomly SNR levels ranging from -5dB to 10dB. Noise signals are from the Demand dataset \cite{ref26}, along with the clean speech recordings are used to create the noisy speech for the training and validation set. 

Test set contains 500 utterances under the SNR condition of -5dB, 0dB, 5dB and 10dB. Both speakers and noise conditions in the test set are totally unseen by the training set.  All of the clean speech and noise recordings with a sampling frequency of 16 kHz, and the frame size and frame shift for STFT are set to 512 and 256.

\subsection{Training and network parameters}
In each epoch of the training, we chunk a random segment of 4 seconds from an utterance if it is larger than 4 seconds. The smaller utterances are zero-padded to match the size of the largest utterance in the batch. The Adam optimizer is used for stochastic gradient descent (SGD) based optimization. The numbers of output channels for the layers in the encoder are changed to 16, 32, 64, 128 and 256 successively, and those for each layer in the decoder to 128, 64, 32, 16 and 1 successively. The initial learning rate is set to 0.001, and is decreased by 50$\%$ once when consecutive 3 validation stagnates, and the training is early-stopped when 10 consecutive validation loss increments happened.

All training audio data are downsampled to 16 kHz and features are extracted by using frames of length 512 with frame shift of 256, and Hann windowing followed by STFT of size $K=512$ with respective zero-padding. Real and imaginary parts of the complex spectrum are divided into separate feature maps, resulting in $C=2$ input channels, thus the input. In addition, the hyper-parameter $\alpha$ of time-frequency domain loss in Eq.21 is set to 0.8 \cite{ref27}.

\renewcommand{\arraystretch}{0.95}
\begin{table*}[]
\LARGE
\caption{Comparisons of Alternative Models in STOI, PESQ, and SNR.}
\resizebox{\textwidth}{!}{
\begin{tabular}{|c|ccc|ccc|ccc|ccc|c|}
\hline
Test SNR                & \multicolumn{3}{c|}{-5dB}                      & \multicolumn{3}{c|}{0dB}                        & \multicolumn{3}{c|}{5dB}                        & \multicolumn{3}{c|}{10dB}                       & \multirow{2}{*}{\# Param.} \\ \cline{1-13}
Metric                  & STOI(\%)       & PESQ          & SSNR(dB)       & STOI(\%)       & PESQ          & SSNR(dB)        & STOI(\%)       & PESQ          & SSNR(dB)        & STOI(\%)       & PESQ          & SSNR(dB)        &                            \\ \hline
Unprocessed             & 57.71          & 1.37          & -5.04         & 71.02          & 1.73          & -0.05          & 82.53          & 2.03          & 4.93           & 90.41          & 2.42          & 9.79           & -                          \\ \hline
OMLSA \cite{ref28}                   & 61.23          & 1.54          & 2.20          & 75.43          & 2.06          & 6.10           & 86.27          & 2.31          & 8.90           & 92.66          & 2.68          & 12.80          & -                          \\ \hline
Attention-wave-U-net \cite{ref29}                     & 78.81          & 2.16          & 6.54          & 86.76          & 2.41          & 8.21           & 92.43          & 2.87          & 10.13          & 94.07          & 3.09          & 15.49          & 10.31 M                     \\
Attention-DCUNet \cite{ref30}            & 81.45          & 2.23          & 6.64          & 88.76          & 2.54          & 8.35           & 92.96          & 2.96          & 11.01          & 95.74          & 3.17          & 15.92          & 2.36 M                     \\ 
TSTNN \cite{ref31}               & 83.50          & 2.38          & 6.97         & 89.54          & 2.66          & 8.74           & 93.29          & 3.04          & 11.44          & 96.08          & 3.26          & 16.31          & 1.54 M                     \\ \hline
U-Former (Proposed) & \textbf{85.48}          & \textbf{2.46}          & \textbf{7.24}          & \textbf{91.69 }         & \textbf{2.78}          & \textbf{9.47}          & \textbf{94.16}          & \textbf{3.13}          & \textbf{12.29}          & \textbf{96.89}          & \textbf{3.35}          & \textbf{16.60}          & 2.03 M                   \\ \hline
\end{tabular}}
\label{table:1}
\end{table*}

\subsection{Baselines}

The baseline systems used for performance comparison are Optimally Modified Log Spectral Amplitude (OMLSA) method \cite{ref28}, attention-wave-U-net \cite{ref29}, Attention-DCUNet \cite{ref30}, and TSTNN \cite{ref31}. The setup of the baseline systems are detailed as below:
\begin{itemize}
    \item \textbf{OMLSA}: A statistical-based noise reduction algorithm.
    \item \textbf{Attention-wave-U-net}: Attention-wave-U-net is an end-to-end speech enhancement model in the time-domain, which contains 12 trainable blocks and is trained on randomly-sampled audio excerpts using Adam optimizer, with learning rate = $10^{-4}$ and batch size of 16. We use $F = 24$ filters for convolution in first layer of network, downsampling block filters of size 15 and upsampling block filters of size 5 as in \cite{ref32}.
    \item \textbf{Attention-DCUNet}: Attention-DCUNet takes RI-spectrograms as input and output, and applies joint time-frequency loss to train the model. The attention-DCUnet has 20 convolutional layers and applies STFT with a 64 ms sized Hann window and 16 ms hop length. We set learning rate to $10^{-4}$ and decay it by 0.5 when the validation score increases. 
    \item \textbf{TSTNN}: TSTNN is a time-domain speech enhancement model, which takes s four stacked two-stage Transformer blocks to efficiently extract local and global information from the encoder output stage by stage. The TSTNN is trained for 100 epochs and optimized by Adam. We use the gradient clipping with maximum L2-norm of 5 to avoid gradient explosion. For learning rate, we use the dynamic strategies during the training stage \cite{ref24}.
\end{itemize}

\subsection{Evaluation Metrics}
The following three metrics are used to evaluate M2BSE-Net and state-of-the-art competitors. All metrics are better if higher.
\begin{itemize}
    \item PESQ: Perceptual evaluation of speech quality (from $-0.5 $ to $4.5$) \cite{ref33}.
    \item STOI: Short-time objective intelligibility measure (from $0$ to $100(\%)$) \cite{ref34}.
    \item SSNR: Segmental SNR \cite{ref35}.
\end{itemize}.

\section{Results and Analysis}
\subsection{Model Comparison}

Comprehensive comparison among alternative models are shown Table~\ref{table:1}, in terms of STOI, PESQ, and SSNR, where the numbers represent the averages over the test set in each condition. The best scores in each case are highlighted by boldface.

We firstly compare the OMLSA with other 4 models. As shown in Table~\ref{table:1}, the OMLSA achieves about $2.25\%$ to $4.43\%$ STOI improvements, $0.17$ to $0.33$ PESQ improvements, and $3.01$dB to $7.24$dB SSNR improvements over unprocessed mixtures. Going from OMLSA to attention-wave-U-net substantially improves the 3 metrics. This result is consistent with the finding in \cite{ref36}. Even OMLSA enhance noisy speech by minimizing the difference between the clean speech and the estimated clean speech, which is the most effective algorithm for traditional monaural speech enhancement \cite{ref37}, ], the OMLSA unable to track a target speaker in a complex acoustic environment. In contrast, the other 4 models are shown their reliability of performance and robustness in different noise types and SNR levels.

Unlike the traditional monaural speech enhancement, the two attention based deep learning models (i.e. attention-wave-U-net and attention-DCUNet) formulate the speech enhancement as supervised learning and utilizes attention for improving the network performance. The two deep learning based models, in which the discriminative patterns within speech or noise signals are learned from training data, is more advantageous for modelling target speech. It is worth noting that attention-DCUNet estimates clean RI-spectrograms directly from noisy speech. As shown in Table~\ref{table:1}, although a similar performance is obtained by attention-wave-U-net and attention-DCUNet, attention-DCUNet generalizes better than attention-wave-U-net.

\begin{figure}
\centering
\subfigure[Noisy Spectrum]{
\label{Fig.sub.1}
\includegraphics[width=0.225\textwidth]{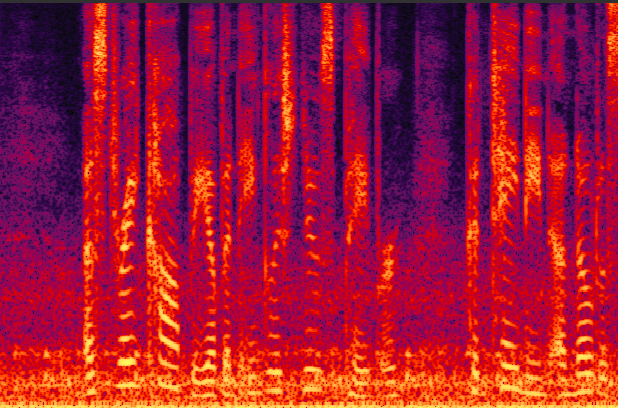}}
\subfigure[Enhanced Spectrum (OMLSA)]{
\label{Fig.sub.2}
\includegraphics[width=0.23\textwidth]{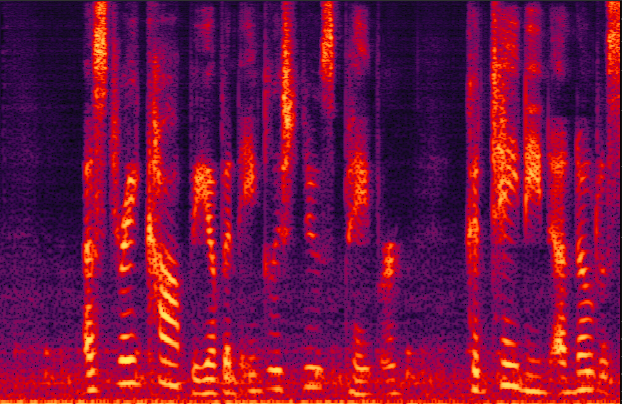}}
\subfigure[Enhanced Spectrum (Attention-wave-u-net)]{
\label{Fig.sub.3}
\includegraphics[width=0.23\textwidth]{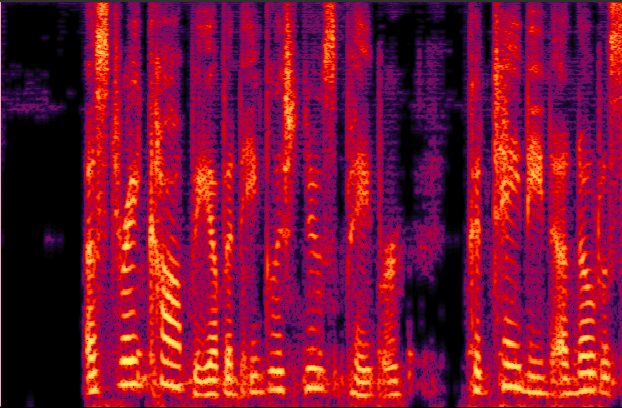}}
\subfigure[Enhanced Spectrum (Attention-DCUNet)]{
\label{Fig.sub.4}
\includegraphics[width=0.23\textwidth]{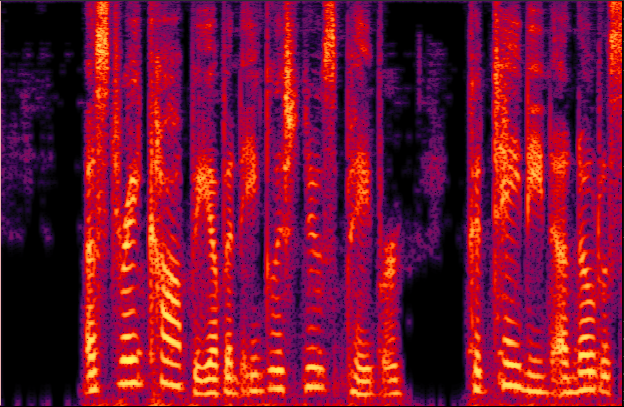}}
\subfigure[Enhanced Spectrum (TSTNN)]{
\label{Fig.sub.5}
\includegraphics[width=0.23\textwidth]{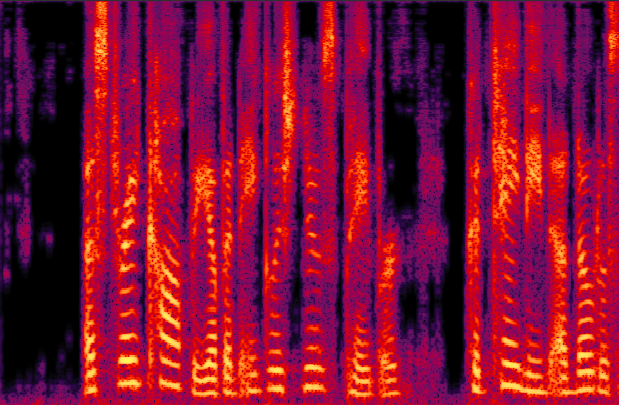}}
\subfigure[Enhanced Spectrum (U-Former)]{
\label{Fig.sub.6}
\includegraphics[width=0.23\textwidth]{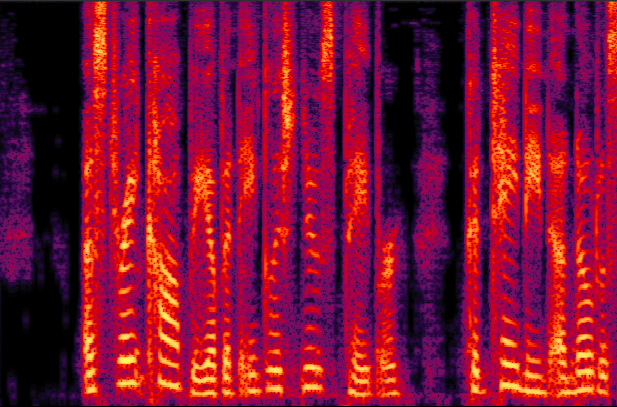}}
\caption{Example of input and enhanced spectrum from an example speech utterance.}
\label{fig.main}
\end{figure}

In addition, the TSTNN is designed to use two stage Transformer block to efficiently extracts both local and global contextual information for long-range speech sequences. Although TSTNN is an end-to-end speech enhancement model in the time domain, it obtains consistently better performance and shows a more robust and better estimation than attention-DCUNet. This further demonstrates the effectiveness of Transformer for tracking a target speaker by leveraging long-term contexts. The proposed U-Former consistently outperforms attention-wave-U-net, attention-DCUNet, and TSTNN in all conditions. Take, for example the $-5$ dB SNR case. The proposed U-Former improves STOI by $1.92\%$, PESQ by $0.08$, SSNR by $0.27$ dB over TSTNN. Figure~\ref{fig.main} shows the visualization of OMLSA, Attention-wave-u-net, Attention-DCUNet, TSTNN, and U-Former enhancement in $0$ dB case.

\renewcommand{\arraystretch}{0.95}
\begin{table*}[]
\caption{Evaluation of MHSA and MHCA for U-Former }
\centering

\begin{tabular}{|c|cc|cc|cc|cc|}
\hline
Test SNR                & \multicolumn{2}{c|}{-5 dB}                          & \multicolumn{2}{c|}{0 dB}                           & \multicolumn{2}{c|}{5 dB}                           & \multicolumn{2}{c|}{10 dB}                          \\ \hline
Metric                  & \multicolumn{1}{c|}{STOI(\%)}       & PESQ          & \multicolumn{1}{c|}{STOI(\%)}       & PESQ          & \multicolumn{1}{c|}{STOI(\%)}       & PESQ          & \multicolumn{1}{c|}{STOI(\%)}       & PESQ          \\ \hline
U-Former-w/o-MHCA\&MHSA & \multicolumn{1}{c|}{74.54}          & 2.03          & \multicolumn{1}{c|}{80.34}          & 2.39          & \multicolumn{1}{c|}{83.35}          & 2.74          & \multicolumn{1}{c|}{86.49}          & 2.98          \\ \hline
U-Former-w/o-MHSA       & \multicolumn{1}{c|}{79.31}          & 2.37          & \multicolumn{1}{c|}{86.79}          & 2.65          & \multicolumn{1}{c|}{93.09}          & 3.06          & \multicolumn{1}{c|}{95.37}          & 3.26          \\ \hline
U-Former-w/o-MHCA       & \multicolumn{1}{c|}{80.45}          & 2.42          & \multicolumn{1}{c|}{86.81}          & 2.68          & \multicolumn{1}{c|}{93.25}          & 3.09          & \multicolumn{1}{c|}{95.98}          & 3.29          \\ \hline
U-Former                & \multicolumn{1}{c|}{\textbf{85.48}} & \textbf{2.46} & \multicolumn{1}{c|}{\textbf{91.69}} & \textbf{2.78} & \multicolumn{1}{c|}{\textbf{94.16}} & \textbf{3.13} & \multicolumn{1}{c|}{\textbf{96.89}} & \textbf{3.35} \\ \hline
\end{tabular}
\label{table:3}
\end{table*}

\renewcommand{\arraystretch}{1.15}
\begin{table*}[]
\centering
\caption{Comparison of different loss functions of U-Former.}
\resizebox{\textwidth}{!}{
\begin{tabular}{|c|cccc|ccc|ccc|ccc|}
\hline
Test SNR                      & \multicolumn{4}{c|}{-5 dB}                                                                                          & \multicolumn{3}{c|}{0 dB}                                                                & \multicolumn{3}{c|}{5 dB}                                                                 & \multicolumn{3}{c|}{10 dB}                                                                \\ \hline
Metric                        & \multicolumn{1}{c|}{}    & \multicolumn{1}{c|}{STOI (\%)}      & \multicolumn{1}{c|}{PESQ}          & SSNR (dB)      & \multicolumn{1}{c|}{STOI (\%)}      & \multicolumn{1}{c|}{PESQ}          & SSNR (dB)      & \multicolumn{1}{c|}{STOI (\%)}      & \multicolumn{1}{c|}{PESQ}          & SSNR (dB)       & \multicolumn{1}{c|}{STOI (\%)}      & \multicolumn{1}{c|}{PESQ}          & SSNR (dB)       \\ \hline
Mixture                       & \multicolumn{1}{c|}{}    & \multicolumn{1}{c|}{57.71}          & \multicolumn{1}{c|}{1.37}          & -5.04         & \multicolumn{1}{c|}{71.02}          & \multicolumn{1}{c|}{1.73}          & -0.05         & \multicolumn{1}{c|}{82.53}          & \multicolumn{1}{c|}{2.03}          & 4.93           & \multicolumn{1}{c|}{90.41}          & \multicolumn{1}{c|}{2.42}          & 9.79           \\ \hline
\multirow{2}{*}{U-Former-T}   & \multicolumn{1}{c|}{MAE} & \multicolumn{1}{c|}{83.44}          & \multicolumn{1}{c|}{2.05}          & 6.67          & \multicolumn{1}{c|}{89.94}          & \multicolumn{1}{c|}{2.52}          & 9.14          & \multicolumn{1}{c|}{93.52}          & \multicolumn{1}{c|}{2.90}          & 11.90          & \multicolumn{1}{c|}{95.89}          & \multicolumn{1}{c|}{3.10}          & 16.56          \\ \cline{2-14} 
                              & \multicolumn{1}{c|}{MSE} & \multicolumn{1}{c|}{83.61}          & \multicolumn{1}{c|}{2.14}          & 6.79          & \multicolumn{1}{c|}{90.01}          & \multicolumn{1}{c|}{2.57}          & 9.18          & \multicolumn{1}{c|}{93.61}          & \multicolumn{1}{c|}{3.00}          & 12.03          & \multicolumn{1}{c|}{96.01}          & \multicolumn{1}{c|}{3.13}          & 16.63          \\ \hline
\multirow{2}{*}{U-Former-RI}  & \multicolumn{1}{c|}{MAE} & \multicolumn{1}{c|}{85.27}          & \multicolumn{1}{c|}{2.21}          & \textbf{7.72} & \multicolumn{1}{c|}{91.35}          & \multicolumn{1}{c|}{2.66}          & 9.39          & \multicolumn{1}{c|}{94.13}          & \multicolumn{1}{c|}{2.94}          & \textbf{12.38} & \multicolumn{1}{c|}{96.71}          & \multicolumn{1}{c|}{3.14}          & \textbf{17.12} \\ \cline{2-14} 
                              & \multicolumn{1}{c|}{MSE} & \multicolumn{1}{c|}{83.43}          & \multicolumn{1}{c|}{2.29}          & 7.40          & \multicolumn{1}{c|}{90.06}          & \multicolumn{1}{c|}{2.73}          & 8.94          & \multicolumn{1}{c|}{93.86}          & \multicolumn{1}{c|}{3.05}          & 12.09          & \multicolumn{1}{c|}{96.52}          & \multicolumn{1}{c|}{3.20}          & 17.00          \\ \hline
\multirow{2}{*}{U-Former-TF1} & \multicolumn{1}{c|}{MAE (for time-domain loss)} & \multicolumn{1}{c|}{\textbf{85.48}} & \multicolumn{1}{c|}{\textbf{2.46}} & 7.24          & \multicolumn{1}{c|}{91.69}          & \multicolumn{1}{c|}{2.78}          & \textbf{9.47} & \multicolumn{1}{c|}{\textbf{94.16}} & \multicolumn{1}{c|}{3.13}          & 12.29          & \multicolumn{1}{c|}{\textbf{96.89}} & \multicolumn{1}{c|}{\textbf{3.35}} & 16.60          \\ \cline{2-14} 
                              & \multicolumn{1}{c|}{MSE (for time-domain loss)} & \multicolumn{1}{c|}{83.57}          & \multicolumn{1}{c|}{\textbf{2.46}} & 6.79          & \multicolumn{1}{c|}{90.82}          & \multicolumn{1}{c|}{2.71}          & 9.14          & \multicolumn{1}{c|}{94.00}          & \multicolumn{1}{c|}{3.04}          & 12.17          & \multicolumn{1}{c|}{96.03}          & \multicolumn{1}{c|}{3.13}          & 16.29          \\ \hline
\multirow{2}{*}{U-Former-TF2} & \multicolumn{1}{c|}{MAE (for time-domain loss)} & \multicolumn{1}{c|}{85.36}          & \multicolumn{1}{c|}{2.45}          & 6.96          & \multicolumn{1}{c|}{\textbf{91.75}} & \multicolumn{1}{c|}{\textbf{2.80}} & 9.25          & \multicolumn{1}{c|}{\textbf{94.16}} & \multicolumn{1}{c|}{\textbf{3.15}} & 12.06          & \multicolumn{1}{c|}{96.64}          & \multicolumn{1}{c|}{3.26}          & 16.42          \\ \cline{2-14} 
                              & \multicolumn{1}{c|}{MSE (for time-domain loss)} & \multicolumn{1}{c|}{83.62}          & \multicolumn{1}{c|}{\textbf{2.46}} & 6.63          & \multicolumn{1}{c|}{90.96}          & \multicolumn{1}{c|}{2.64}          & 9.08          & \multicolumn{1}{c|}{93.88}          & \multicolumn{1}{c|}{2.98}          & 11.94          & \multicolumn{1}{c|}{96.15}          & \multicolumn{1}{c|}{3.11}          & 16.38          \\ \hline
\end{tabular}}
\label{table:2}
\end{table*}

\subsection{Evaluation of MHSA and MHCA}
The proposed U-Former takes MHSA and MHCA for leveraging long-term correlations and dependencies and filtering out uncorrelated features. In this study, we remove the MHSA module, MHCA module, and both multi-head attention modules of U-Former respectively, and compare the performance between these three ``incomplete U-Former'' and the ``complete U-Former''.

From From Table~\ref{table:3}, we can conclude that the existence of the MHSA module and the MHCA module does promote the U-Former performance. Take, for example the $0$ dB case, the MHSA and MHCA improve $10.35\%$ and $0.39$ on STOI and PESQ respectively. Furthermore, ``U-Former-w/o-MHCA'' performs better than ``U-Former-w/o-MHSA'' in all SNR cases in terms of STOI and PESQ. Take, for example the $-5$ dB SNR case, the ``U-Former-w/o-MHCA'' improves STOI by $1.14\%$, and PESQ by $0.05$ over ``U-Former-w/o-MHSA''

\subsection{Evaluation of Loss Function}
The proposed U-Former takes joint time-frequency loss as training objective. To explore the effectiveness of joint time-frequency loss, we train the U-Former with different loss functions. All the time domain models are compared for both the MAE loss and the MSE loss training. The corresponding abbreviated names for our models trained using different loss functions are U-Former-T for using loss function on time domain samples, U-Former-T for using a loss function on both thr real and the imaginary part of STFT, and U-Former-TF for using a loss on time domain and STFT magnitudes. In particular, U-Former-TF1 denotes the frequency loss defined on L1 norm and U-Former-TF2 denotes the frequency loss defined on L2 norm. Table~\ref{table:2} shows the average results at $-5$ dB, 0 dB, 5 dB and 10 dB. In addition, the U-Former-TF1 and U-Former-TF2 using MAE and MSE based time-domian loss are compared in this table.

From Table~\ref{table:2}, we can firstly conclude that the existence of the joint time-frequency loss does promote the network performance. What is more, we observe that the models trained with a time-domain loss perform better using MSE. The U-Former-T, U-former-TF1 and U-former-TF2 have better STOI, PESQ and SSNR scores with MAE based time-domain loss. In addition, the models trained with with a frequency-domain loss perform better using MAE. The U-former-RI have better STOI and SSNR scores with MSE based frequency-domian loss but significantly worse PESQ. Finally, U-former-TF1 and U-former-TF2 have similar STOI and PESQ scores but U-former-TF1 is consistently better in terms of SSNR.

\section{Conclusion}\label{sec6}
This paper proposed a U-Former for monaural speech enhancement network with MHSA and MHCA. The model adopts U-Net architecture, which applies multi-head self-attention as bottleneck, multi-head cross-attention as skip-connection, and a novel time-frequency loss. According to the experiment results, both attention mechanisms, and time-frequency loss improve performance of speech enhancement systems by involving U-Net architecture. Through experiments on two datasets, we show the superiority of the proposed method over prior arts.

In the near future, we plan to speed up our model and apply it to low-latency applications. Another important direction is to study how different training targets will influence the proposed method. Finally, we plan to extend the proposed method to various audio-related tasks such as dereverberation.


%
\bibliographystyle{IEEEtran}
\bibliography{IEEEabrv, IEEEexample}

\end{document}